# Universal 2D Soft Nano-Scale Mosaic Structure Theory for Polymers and Colloids

Jia-lin WU

College of Material Science and Engineering, Donghua University, Shanghai 201620, China
E-mail: jlwu@dhu.edu.cn

**ABSTRACT**

A basic concept in chain-particle cluster-motion, from frozen glassy state to melt state, is the 2D soft nano-scale mosaic structure formed by 8 orders of 2D interface excitation (IE) loop-flows, from small to large in inverse cascade and rearrangement structure in cascade along local one direction. IE has additional repulsive energy and extra vacancy volume. IE results from that the instantaneous synchronal polarized electron charge coupling pair is able to parallel transport on the interface between two neighboring chain-particles with antiparallel delocalization. This structure accords with de Gennes' mosaic structure picture, from which we can directly deduce glass transition temperature, melt temperature, free volume fraction, critical entangled chain length, and activation energy to break solid lattice. This is also the inherency maximum order-potential structure in random systems.

## 1. Introduction

This letter is an introduction of the motion principle in polymer physics theory. The confluence of both the thermodynamic and the kinetic dimensions of the solid ↔liquid glass transition (GT) presents one of the most formidable problems in condensed matter physics [1]. Many theories [2], containing current Mode-Coupling Theories, have been proposed to explain glasses and GT; however, GT theory is still an open problem [2]. An acceptable GT theory should be in accordance with de Gennes' simple mosaic structure picture [3]. (a) The size of the clusters at GT corresponds to the boson peak wavelength, clusters slightly more compact than the matrix: they cannot grow in size because of frustration effects. Mode-Coupling Theories cannot describe this, because they do not incorporate frustration. (b) Clusters move rather than molecules, the required cavity space ('vacancy') for cluster motion is not empty, but filled with the low density matrix. (c) The soft matrix is surrounding the clusters. (d) An unsolved problem is what mechanism to balance kinetic and potential at the GT. In polymer engineering, GT may be thought of as an inverse cascade – cascade mode along external stress direction. This idea comes from the insight for cooperative orientation activation energy, $\Delta E_{co}$, on polyester melt high-speed spinning-line [4]. When the work of the stress on more 5000M/min spinning-line reaches $\Delta E_{co}$, the structure of the yarn is stable and reaches full orientation, called as FOY (Full Orientation Yarn) in current polyester fiber industry. This phenomenon is called stress-induced GT. The rate of change of the stress-induced liquid-to-solid GT is $10^7$ times of that in general quencher from melt state to frozen glass state. A logical explanation is that the macromolecules can complete liquid-to- -solid GT with full orientation within the millisecond of time in $z$-space. Their motion mode, within the entire range from melt transition temperature $T_m$ to GT temperature $T_g$, allows the direction of inverse cascade – cascade in every "excited domain" is in arbitrary in melt state and all apt to $z$-axial on melt spinning-line; and finally the mode is frozen in glasses as the soft matrix.

On the other hand, in physical theory, many complicated phenomena originate from the global properties of parallel transport of simple quantum systems (Berry's Phase) [5]. In addition, crucial to the endeavor of GT theory is a deeper understanding of the *systematics of bonding* in condensed matter within a framework going considerable beyond the current GT picture [6]. Thus, the GT theoretical approach may come down to the *parallel transport of "bonding"* on intermolecular interfaces. This "bonding" here is called as the *interface excitation* (IE) the new and crucial concept introduced in this letter. Nine physical ingredients, random, self-similar, two- -body interaction, fluctuation, frustration, and percolation, delocalization, Berry's phase and

Brownian regression potential, have been incorporated at the GT and in macromolecular motion. The incorporation of physical ingredients results in the 2D soft nano-scale mosaic structure in "excited domain".

## 2. Theoretical Model

### 2.1. Intermolecular Interface Excitation

Van der Waals interaction includes the contribution of instantaneous induced dipole – induced dipole. Generally, instantaneous polarized dipole electron charges randomly distribute on an interface 1-2 forming electron cloud (blue zone) on x-y projection plane in **Figure 1(a)**.

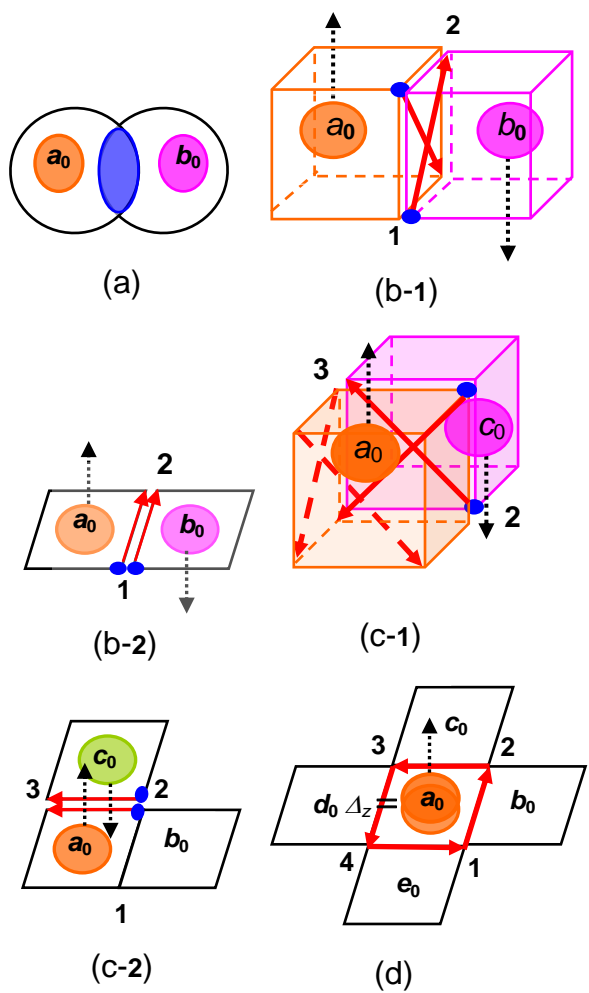

**Figure 1. The physical origin of interface excitation. (a) A legitimate state. (b), (c) An absent of attraction state in certain 2D lattices. (d) Instantaneous delocalizing state of ion.**

At GT, an interesting and unexplored corner of Van der Waals interaction theories is that the instantaneous synchronal polarized electron charge coupling pair (two small blue dots) may *parallel transport* on an interface, e.g., the interface 1-2 between particles $a_0$ and $b_0$ in **Figure 1(b)** during the local time of $(t_{0,0}, t_{0,1})$, and the interface 2-3 between $a_0$ and $c_0$ in **Figure 1(c)** during the local time of $(t_{0,1}, t_{0,2})$. The state of parallel transport of coupling pair is defined as interface excitation on 2D projection plane. Black-broken line arrow on an ion denotes its delocalizing direction in **Figure 1**.

Thus, the site-phase difference between two *z*-component molecules, or chain-particles, is $\pi$, the state of all polarized electron charges in each instantaneous dipole in the two *z*-component chain-particles, in **Figure 1 (b-1)** or **(c-1)**, must be in the *same state* and in the *z*-axial *minimum energy state* of single chain-particles instantaneous dipole. In **Figure 1**, **(b-2)** and **(c-2)** is respectively the projection of **(b-1)** and **(c-1)** on x-y plane. This means IE has an additional repulsive energy $\Delta\varepsilon$ and an extra vacancy volume $\Delta v$. Two instantaneous synchrony *z*-axial polarized electron charges parallel transport from one end to other end on an interface, as in **Figure 1(b-2)**, that is simply denoted by an arrowhead, as (1→2) in **Figure 1(d)**. The two IE states of **(b-1)** and **(c-1)** occur at different local times. The directions of next two parallel transports of instantaneous polarized electron charges of the reference $a_0$ particle denote as two red-broken line arrowheads in **Figure 1(c-1)**. Thus, an IE loop-flow 1→2 → 3 → 4→ 1, occurs during the local time of $(t_{0,0}, t_{0,4})$ denoted as loop-$V_0(a_0)$, in **Figure 1(d)**. It is able to offer a non-integrable Berry's Phase potential $\Delta\varepsilon$ to induce ion $a_0$ (the center of mass of $a_0$ particle) *z*-direction a conceivable displacement $\Delta_z$.

### 2.2. 5-Particle Cooperative Excited Field

The IE loop-flow in **Figure 1(d)** also defined a 2D action particle $a_0$ to GT, also denoted as $V_0(a_0)$. The needful time $(t_{0,0}, t_{0,4})$ to form loop-$V_0(a_0)$ defined as the zero order of relaxation time $\tau_0$. Thus each IE **in Figure 1** has also the relaxation time of $\tau_0$ scale. At the GT, because only 2D closed cycle-flow (that will generate a non-integrable phase $\pi$ and an additional *z*-axial induced energy) corresponds to the abnormal heat capacity and boson peak [3, 7–9], we need only discuss all the IE loop-flows on reference $a_0$ (particle local excited) filed. The discussion further simplify to only consider the loop-flows formed by several consecutive arrows on a 2D projection plane, see **Figures 2**–**5**, and each arrow denotes an IE that has the same IE energy $\Delta\varepsilon$ for flexible chain system. The predominance of the model is also able to describe complex system. Some arrows with different markers denote the different IE energies correspond to the non-flexible chain system.

The formation of a central excited particle $V_0(a_0)$ can be also regarded as the cooperative contribution from its 4 neighboring particle fields, $V_0(b_0)$, $V_0(c_0)$, $V_0(d_0)$, and $V_0(e_0)$ based on Brownian motion theory **Figure 2**.

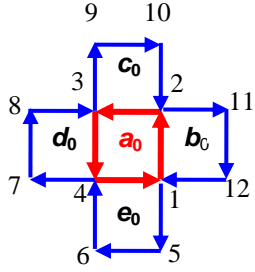

**Figure 2.** First order of transient 2D IE loop-flow $V_1(a_0)$.

At the instant (local) time $t_1$ in $a_0$ field, once 4th loop-flow $V_0(e_0)$ is finished, a new loop-flow of $a_0$ surrounded by the 12 IEs, 1→ 5→ 6→ 4→ 7→ 8→ 3→ 9→ 10→ 2→ 11→ 12→1, with $\tau_1$ timescale will occur as in **Figure 2**. The new symmetric loop-flow defined as the *first order of transient 2D IE loop-flow* and *first order of 2D cluster* in $a_0$ field, which are all denoted as $V_1(a_0)$. Whose cycle direction is negative, contrary to that of $V_0(a_0)$. The energy of IE with $\tau_1$ on the loop denoted as $\Delta\varepsilon(\tau_1)$. The 4 excited particles are 4 concomitant mosaic cells with $V_1(a_0)$. Loop $V_1(a_0)$ has 12 IEs and is of the loop potential energy of $12\Delta\varepsilon_0(\tau_1)$. This means that the evolution energy from $V_0(a_0)$ to $V_1(a_0)$ is $8\Delta\varepsilon(\tau_1)$. The number of cooperative excitation particles in $V_1(a_0)$ is 5.

Similarly, the 2nd order of transient 2D IE loop-flow and 2D cluster with $\tau_2$, denoted as $V_2(a_0)$, is formed at the instant local time $t_2$ when its 4 neighboring particle fields: $V_1(c_0)$, $V_1(d_0)$, $V_1(e_0)$, and $V_1(b_0)$ finished in $a_0$ field. Whose cycle direction is positive, contrary to that of $V_1(a_0)$, in **Figure 3 (e)**. The yellow area in **Figure 3(e)** represents $V_1(c_0)$ loop in $c_0$ field, which is a fist order of mosaic structure in $V_2(a_0)$. The streamline diagram of parallel transport between two clusters $V_1(c_0)$ and $V_1(a_0)$ is **Figure 3(a)**. 4 such mosaic structures, $V_1(c_0)$, $V_1(d_0)$, $V_1(e_0)$, and $V_1(b_0)$ cooperatively generate $V_2(a_0)$ in $a_0$ field.

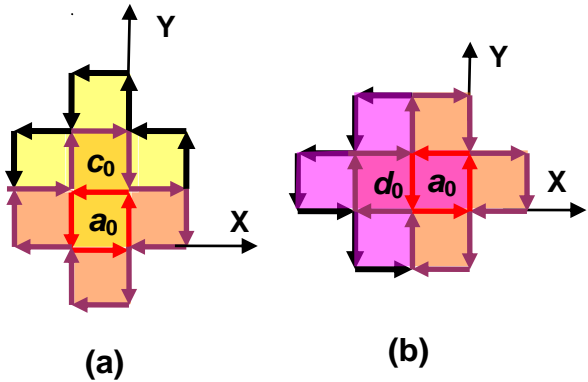

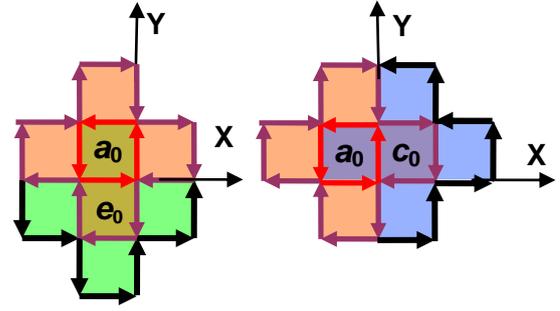

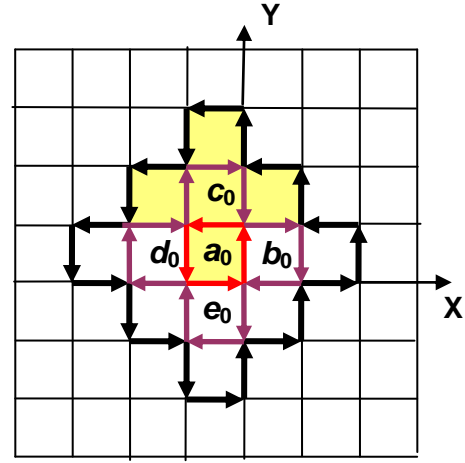

**Figure 3.** 4 neighboring clusters, $V_1(c_0)$, $V_1(d_0)$, $V_1(e_0)$ and $V_1(b_0)$ respectively and widdershins interact with $V_1(a_0)$: (a), (b), (c), and (d) cooperatively generate $V_2(a_0)$, (e). The 20 thick arrows denote 2nd order of 2D loop-flow and cluster $V_2(a_0)$ in $a_0$ field.

The energy of IE on loop $V_2(a_0)$ is denoted as $\Delta\varepsilon(\tau_2)$. The number of IEs of loop $V_2(a_0)$ is 20. The number of IEs inside loop $V_2(a_0)$ is 12, and equals to the number of the IEs of $V_1(a_0)$. This means that the evolution energy from $V_1(a_0)$ to $V_2(a_0)$ is $8\Delta\varepsilon(\tau_2)$. The number of cooperative excitation particles in $V_2(a_0)$ is 13 surrounded by 20 IEs. Thus, 20 IEs can excite 13 particles to hop randomly along local $+z$-axial direction.

In the same way, the third order of 2D IE loop-flow and cluster at the instant time $t_3$ in $a_0$ field, $V_3(a_0)$, can be obtained as in **Figure 4**, which formed by the contribution of its 4 neighboring second order loop-flows, $V_2(e_0)$, $V_1(b_0)$, $V_1(c_0)$, and $V_1(d_0)$, respectively and deasil interacted with $V_2(a_0)$ in $a_0$ field. In **Figure 4**, the yellow area represents loop $V_2(e_0)$ in $e_0$ field (whose symmetric center is $e_0$) which is a second order of mosaic structure in $a_0$ field. The number of IEs of loop $V_3(a_0)$ is 28. The number of IEs inside loop $V_3(a_0)$ is 20, which equals to the number of the IEs on loop $V_2(a_0)$. This means the evolution energy from $V_2(a_0)$ to $V_3(a_0)$ is 8 $\Delta\varepsilon(\tau_3)$. The number of cooperative excitation particles in third order cluster is 25. Thus, 28 IEs can excite 25 particles to hop randomly along local $-z$-axial direction.

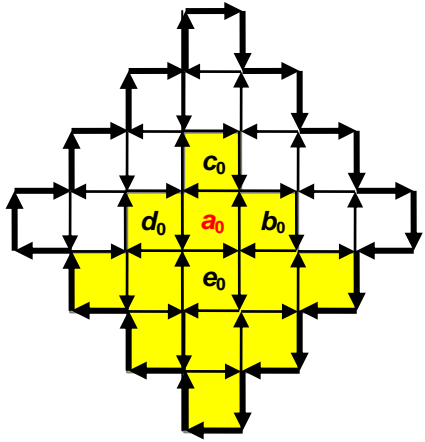

**Figure 4**. Third order of transient 2D IE loop-flow and cluster $V_3(a_0)$ in $a_0$ field.

From the results of $V_1(a_0)$, $V_2(a_0)$ and $V_3(a_0)$, it is clear that the more the number of cooperatively delocalization particles is, the less the average needful IE energy for each particle is. The key point is what the minimum excited energy is and what the number of cooperatively delocalization particles excited by the energy at the GT is.

There are 4 neighboring concomitant excitation centers surrounding the referenced particle center. Thus, the excited field in **Figure 2** is defined as *5-particle cooperative excited field to break solid-lattice*.

The forming of *i*-th order cluster $V_i(a_0)$ always results from the cooperative contributions of its 4 neighboring (*i*-1)-th order of clusters around the central particle. This property origins from Brownian motion and is called as the *priority* of central particle, which is probably the primary reason to generate dynamic heterogeneity [2] and non-ergodic [9] at GT. That means the spatial directions and localities of all the succedent IEs will be completely confirmed, **Figures 2 – 5**, once the *first* IE with the direction 1 → 2 showed in **Figure 1(d)** appears at the local time $t_{0,1}$ in $a_0$ field, otherwise, the total excitation energies will increase. That is the reason and advantage to bring the signature of arrowhead into the standout mode of IE to get the minimum energy required to excite GT. Here the IE represented by arrow is scaled by the IE energy $\Delta\varepsilon(\tau_i)$ and thus the arrow is of a *unit length* for different wave lengths or harmonic and *anharmonic frequencies* on the (energy) lattice model. IE energy $\Delta\varepsilon$ does not depend on temperature in flexible polymer system.

## 2.3. Eighth Order of 2D Cluster and Percolation

(a) The number of IEs of *i*-th order of 2D cluster can be calculated as $L_i = 4(2i+1) = 12, 20, 28, 36, 44, 52, 60$, and (68) for $i = 1, 2… 8$ and $L_0 = 4$. Note that $L_8$ will be corrected as 60 by geometric frustration-percolation transition.

(b) The number of excitation particles in *i*-th order of cluster is respectively: $N_i = N_{i-1} + 4i = 5, 13, 25, 41, 61, 85, 113$, and (145) for $i = 1, 2… 8$ and $N_0 = 1$. Note that $N_8$ also will be corrected as 136 by geometric frustration-percolation transition.

(c) For *z*-axial *i*-th order 2D cluster, the evolution energy from $V_{i-1}$ to $V_i$ is 8 $\Delta\varepsilon(\tau_i)$, we emphasize, this evolution is orienting. The energy to excite *i*-th order cluster orientable evolution thus is the energy of *one external degree of freedom* (DoF) to *i*-th order of cluster (loop- -flow) *orienting*, denoted as $\varepsilon_0(\tau_i)$. So we directly deduce from (a)

$$\varepsilon_0(\tau_i) = 8\Delta\varepsilon(\tau_i) \quad (1)$$

**Figure 5.** The 2D soft nano-scale mosaic structure occurs in the geometric frustration – percolation transition and is the maximum order potential structure in random systems.

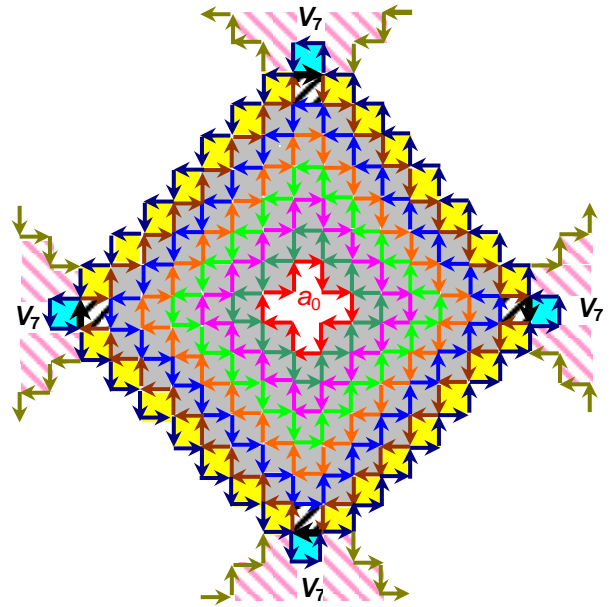

(d) During the time of $(t_{i-1}, t_i)$ in the referenced $a_0$ field, each of its 4 neighboring ($b_0$, $c_0$, $d_0$, and $e_0$) local coordinates can take any direction (fragmentized and atactic lattices) because of fluctuation. However, from (c), *at the instant time $t_i$*, the direction of *i*-th order cluster always starts *sticking to the direction* of the referenced first or-der cluster. This characteristic in random systems comes from the *Brownian regression motion*, or the characteristic of Graph of Brownian Motion. This means that each neighboring *i*-th order of cluster also has one external DoF of energy as $\varepsilon_0(\tau_i)$. So the *i*-th order of cluster in the (*i*+1)-th order is of 5 *inner* DoF and taking any directions during the time of ($t_i$, $t_{i+1}$) in $a_0$ field.

In the same way, the 8th order of transient 2D cluster, $V_8(a_0)$, can be obtained in $a_0$ field, as in **Figure 5**. As there is no 9th cluster in system, (the result of limited domain-wall vibration frequencies based on Random First-Order Transition Theory proposed by Wolynes et al [2, 7]) percolation should be adopted at the instant time $t_8$, and the 8th order of cluster must be corrected by percolation.

The uncorrected number of IEs of $V_8(a_0)$ is 68, denoted by blue-black arrows in **Figure 5**. When percolation appears in system, the 4 excited cells, by thick upward diagonal lines, with $\tau_8$ in $V_8(a_0)$ are also the mosaic cells of in the 4 neighboring $V_7$-clusters (Note: here the 4 $V_7$-clusters are respectively in the 4 neighboring 5-particle cooperative excited fields) with $a_0$ field. And 4 cells with $\tau = \tau_7$ in the 4 neighboring $V_7$-clusters, by the color of blue green, are also the mosaic cells in $V_8(a_0)$. (A region of space that can be identified by a single mean field solution is called a mosaic cell [10]. Mosaic cell is identified by the different action time in our work.) The 4 thick-black *inverted arrows* in **Figure 5** are the *mutual interfaces* of $V_8(a_0)$ and its 4 $V_7$-clusters in neighboring 4 5-particle cooperative excited fields. Thus the number of interfaces of $V_8(a_0)$ should be corrected as 60, that is, $L_8$ (corrected by percolation) = 60.

Note that there are 4 *inverted* thick-black arrows in the corrected $V_8$ loop-flow in **Figure 5**. Thus, the number of excited particles in $V_8(a_0)$ should be firstly corrected as 141 because the 4 mosaic cells of $V_8$, marked by the color of blue green, belong to that of the $V_7$-clusters neighboring 5-particle cooperative excited fields. In addition, the central 5 empty cells, according with de Gennes' simple picture of a localized 'vacancy' among clusters [3], representing the 5 particles in $V_1(a_0)$ have first delocalized (indicates $a_0$ local coordinates invalidation partly!). Therefore, the number of cooperative excited particles in 8th order of cluster should finally be corrected as 136.

## 3. Results and Discussions

### 3.1. Average Energy of Cooperative Migration

In **Figure 5**, each (small square) chain-particle has a *z*-axial hopping energy $\Delta\varepsilon$ that comes from the non-integrable phase of parallel transport of its 4 IEs surround its *z*-axis. From (1), the energy of cooperative migration along z-direction in a domain is $136\Delta\varepsilon = 17\varepsilon_0$. As the migrating direction in different domain can be statistically selected as *x*-, *y*- *z*-axial, and the appearing of all thawing domains is *one by one* and forming flow-percolation, the *average energy of cooperative migration*, $E_{mig}$, *along one-dimensional direction* can be denoted by $kT_2°$, *the non-integrable random Brownian directional regression energy* for *excited particles* inside 8th order of loop-flow. $kT_2°$ can be also balanced by the random thermal motion energy of $kT_2$ for *general unexcited particles*.

$$E_{mig} = kT_2° = kT_2 = 17/3\varepsilon_0 \qquad (2)$$

Here $kT_2$ is similar to the energy of Curie temperature in magnetism, also the energy of a 'critical temperature' existing in the GT presumed by Gibbs based on thermodynamics years ago [11]. The denotation of $kT_2$ is the same as that of Gibbs. $E_{mig}$ along one direction, e.g., along the direction of external stress at GT, is an *intrinsic attractive potential energy, independent of temperature and external stress and response time*. This will be one of the key concepts directly prove the WLF equation. It stands to reason that the attractive potential comes of the IE loop-flows in Brownian motion existing in any random system at any temperature.

### 3.2. Fractal Dimension

The 8 orders of loop-flows in **Figure 5** are the *track records* of selected IEs, from many dipole charge-electron pairs in fluctuation in 3D local space, respectively at the instant of $t_0$, $t_1$, $t_2$…$t_8$ discrete time on a reference $a_0$ particle 2D projection plane. In other words, in 3D $a_0$ local space, the cyclic direction, selected from many random oriented IE loops, returns to the direction of ± *z*-axial at the instant of $t_i$ ($i$ = 0, 1, 2, …8) time, which indicates that the distribution of IE loop in 3D local space obeys the distribution of the Graph of Brownian Motion [12]. Therefore, it can be directly obtain the Hausdorff fractal dimension $d_h$, or the Box fractal dimension $d_c$ of dipole charge electron coupling pairs at the GT is the fractal dimension $d_f$ of Graph of Brownian Motion [12],

$$d_f = d_h = d_c = 3/2 \qquad (3)$$

### 3.3. Eight-Order of Hard-Spheres and Chain Segments

We introduce the concept of *i*-th order of *directional 3D hard-sphere* (hard-cluster) for directional *i*-th order of 2D cluster in order to refract the number of particles in an *orienting and compacted cluster* in density fluctuation when the maximum loop-$V_8(a_0)$ occurs.

(a) **Compacting cluster and density fluctuation.** The first order of hard-sphere $\sigma_1$ contains 5 (the number of particles in $V_1$) + 12 (the number of interfaces of $V_1$, each interface corresponds to an excited particle taking in *any*

*direction* in 3D local space) = 17 chain-particles that are *compacted* to form a hard-sphere. Which has an *internal density larger than average*, because the IEs with extra volumes on $V_0(a_0)$ inside $V_1(a_0)$ have been compacted and transferred to that on $V_1(a_0)$ (static electricity screen effect of IE loop-flow).

**(b) Hard-sphere with finite acting facets.** Note that $\sigma_1$ is a vector. The direction of $\sigma_1$ is negative, as same as the direction of $V_1(a_0)$, if the direction of $V_0(a_0)$ is positive. Finite acting facets surround this orienting 3D hard-sphere. The complexity of dynamical hard-sphere here comes down to the finite acting facets in mosaic structures.

**(c) 8 orders of self-similar z-axial 3D hard-spheres.** In the same way, the second order of hard-sphere $\sigma_2$ contains 13 (the number of particles in $V_2$) + 20 (the number of interfaces of $V_2$, the number of side-excited particles) = 33 excited particles. The direction of $\sigma_2$ is positive. Therefore, the number, $S_i$, of particles in $i$-th order of hard-sphere $\sigma_i$ can be obtained as

$$S_i = -17, +33, -53, +77, -105, +137, -173, +200, \quad (6)$$
$$\text{for } i = 1, 2 \ldots 8. \quad (4)$$

In (4), the sign denotes the moving direction along $z$-axial of $i$-th order hard-sphere. By $i$-th order of hard-sphere is meant that the interaction-interface energy of its two-body is the IE energy or transferred energy, $\Delta\varepsilon(\tau_i)$, independent of the distance of two excited cents of two-body in $z$-space.

**(d) 8 orders of self-similar chain-segments.** In **Figure 5**, each of 200 acting (excited) particles connects with a $z$-component chain-particle in a chain. Accordingly, in macromolecular system, the lengths of the 8 orders of $z$-component statistical chain-segments $l_i$ (if chain-long $N \geq 200$) are also categorize as 8 orders

$$l_i = 17, 33, 53, 77, 105, 137, 173, 200;$$
$$\text{for } i = 1, 2 \ldots 8 \quad (5)$$

**(e) Number of structure rearrangements.** The number of particles in the 8th order of hard-sphere, $N(\sigma_8)$, or the 8th order of chain-segment sizes, $l_8$, also is the number of particles $N_c$ in structure rearrangement in $N(\sigma_8)$ and $l_8$, can be easily found out from **Figure 5**. The number of excited particles in corrected $V_8$ is 136. The number of interfaces of $V_8$ before percolation is 68, from which subtracts 4 mutual interfaces (that belong to 4 neighboring local fields) of $V_8$ and its 4 neighboring $V_7$-clusters when percolation appears, and each of the 64 (= 68 − 4) interfaces *respectively* relates to a side-excited particle taking any direction. Thus, the number of acting particles in $\sigma_8$ is 136 + 64 = 200, a *magic number* in mosaic nano-structure, $N(\sigma_8) = l_8 = N_c = 200$.

The numerical value is consistent with the conjectural results of encompassing rearrangements [13]. $N_c$ is also the critical entangled macromolecule length. The numerical value is in accordance with the experimentally determined critical entanglement chain length of ~ 200 [14].

**(f) Evolution direction of cluster growth transition.** From $N_c$ = 200, $\sigma_8 \approx 5.8$ (chain-particle units), which is the size of 'cage' at GT. According to the definitions of 2D clusters and 3D hard-sphere, an interaction of two $z$-space $i$-th order of 3D hard-spheres is always equal to the IE interaction with $\tau_i$. This means that the cluster in [3] turns out to be $i$-th order of hard-sphere. During the time of ($t_i$, $t_{i+1}$), $i$-th order of 2D clusters cannot be welded together, as same as [3], however, at the instant time $t_{i+1}$, all $i$-th order of 2D clusters in local field will be compacted together to form a compacted ($i$+1)-th order of 2D cluster with the evolution direction of $\sigma_1$ in inverse cascade.

### 3.4. Localization-Delocalization (Percolation) Transition Energy, $E_c$

The steady excited energy is exactly the flow-percolation energy in percolation on a continuum of classical model in condensed matter physics. The energy of steadily 'excited state energy flow' in the process of $V_8$ vanishing and reoccurring is defined as the localization - delocalization (percolation) transition energy, named as $E_c$ (the same denotation of $E_c$ as Zallen [1] did) at GT. The 8 orders of mosaic structures directly manifest that the external DoF of an (inverse cascade) excited state energy flow is 1 and the ($i$−1) order inner DoF of a (cascade) $i$-th order cluster in *renewed cluster state* (rearrangement structures) is 5 at GT.

**(a) Macroscopic melt temperature $T_m$.** It is a very important consequence that the external DoF of an energy flow is 5 in the melting state phase transition; and the renewed energy of a microscopic $i$-th order cluster ($i < 8$), $kT_m°(\tau_i)$, being in renewed cluster within ($i$+1)-th order cluster zone, is *numerically* equal to the energy of the macroscopic melting state of $kT_m$. Namely, $kT_m°(\tau_i) = kT_m(\tau_i) = E_c(\tau_i) + \varepsilon_0(b_0, \tau_i) + \varepsilon_0(c_0, \tau_i) + \varepsilon_0(d_0, \tau_i) + \varepsilon_0(e_0, \tau_i)$, $i < 8$, corresponds to 5 inner DoF, in which $E_c(\tau_i) = kT_g(\tau_8)$, (10); $\varepsilon_0(c_0, \tau_i)$ denotes the energy of one external DoF of $i$-th order loop-flow in $c_0$ field. The energy of macroscopic melting state can be denoted as $kT_m(\tau_8)$,

$$kT_m(\tau_8) = kT_g(\tau_8) + 4\varepsilon_0(\tau_8) = kT_m \text{ (for flexible system)}$$

$$kT_m(\tau_8) = kT_g(\tau_8) + \varepsilon_0(b_0, \tau_8) + \varepsilon_0(c_0, \tau_8) + \varepsilon_0(d_0, \tau_8) +$$
$$\varepsilon_0(e_0, \tau_8) = kT_m \text{ (for general system)} \quad (6)$$

**(b) Percolation transition energy $E_c$.** The step (interface) number of $V_8$-loop is $L_8 = 60$. $E_c$, is less than $60\Delta\varepsilon(\tau_8)$ because of the regression state energy effect of IE and the dynamical mosaic structures of IE loop-flows.

The IE energies on $V_8$-loop in **Figure 5** are *shared by* $V_0(\tau_0)$-$V_8(\tau_8)$ interfaces and $V_7(\tau_7)$-$V_8(\tau_8)$ interfaces.

The key concepts are that a few of interfaces, named as $L_{inver}$, on the $V_8$-loop will be excited by the interfaces with relaxation time of $\tau_i < \tau_8$ in new local fields after $a_0$. This is the effect of mosaic structures, or, $L_{inver}$ is the step number of mosaic in slow inverse cascade. The others, named as $L_{cas}$, on the $V_8$-loop will one by one vanish and their excited energy will rebuild many new $\tau_0$-interfaces in the new local fields in fast cascade.

Thus, in the flow-percolation on a continuum, the 60 interfaces of a reference $V_8$-loop in **Figure 5** are *dynamically* divided into two parts: the $L_{cas}$ interfaces that occur at the local time of $t_8$ in $a_0$ field and the $L_{inver}$ interfaces that are the mosaic structures of energy flow and occur at a time after $t_8$. Formula (7) is obtained

$$L_{cas} = L_8 - L_{inver} \tag{7}$$

The energy of $L_{cas}$ is also the fast-process cascade energy of rebuilding new $V_0$ loop-flows when the $L_{cas}$ interfaces vanish. The energy of $L_{inver}$ is the slow-process inverse cascade energy of all $V_i$ loop-flows from $V_0$ to $V_7$ in new local fields.

The balance excited energy of inverse cascade and cascade in flow-percolation (contained many local fields) is exactly the localized energy $E_c$. Therefore, in fast-process cascade, $E_c/\Delta\varepsilon(\tau_8) = L_{cas}$, in which $\tau_8$ is used as timescale of $V_8$ vanishing, $E_c$ here is the cascade potential energy in flow-percolation.

Since the excited energy in inverse cascade is not dissipated, the evolution energy of each order closed loop is $\varepsilon_0(\tau_i) = 8\Delta\varepsilon(\tau_i)$, (1), namely, the 'singular point energy' of *any i*-th order IE closed cycle (Gauss theorem).

Each mosaic step (either the inverted arrow, or the shared interface by $V_7$-$V_8$) in **Figure 5** connects a mosaic closed cycle $V_7(\tau_7)$ that does not belong to the reference $a_0$ local field. Thus, in the slow-process inverse cascade from $V_0$ to $V_7$ in flow-percolation, the number of the inverse cascade energy $E_c$ forming mosaic step is $E_c/\varepsilon_0(\tau_7) = L_{inver}$. For flexible polymer, $\varepsilon_0(\tau_i) = \varepsilon_0(\tau_8) = \varepsilon_0$. Therefore, formula (8) is obtained

$$E_c = 60\Delta\varepsilon(\tau_8) - E_c/8 \tag{8}$$

Or $\qquad E_c(\tau_i) = 20/3\,\varepsilon_0(\tau_i) \tag{9}$

**Eq.8** is a representative mean field formula. It can be seen that the physics meaning of the right term containing left term on equation is the contribution of mosaic structure. **Eq.9** denotes that the delocalized energy $E_c$ (the localized energy, the percolation transfer energy, the transfer energy from inverse cascade to cascade) at the GT has 8 components, $E_c(\tau_i)$, $i = 1, 2…8$ and the numerical value of each component energy is the same, namely, $20/3\,\varepsilon_0$. This is also one of the singularities at GT. $E_c$ is a characteristic invariable with 8 order of relaxation times at the GT, independent of GT temperature $T_g$.

**(c) Glass transition temperature $T_g$.** We see $E_c$ also comes from the directional non-integrable additional energy at GT, instead of general thermo-random motion energy of molecules. Similar to (2), z-axial non-integrable *random regression vibration* energy, $kT_g°(\tau_i)$, of $i$-th order clusters with $\tau_i$ can be used to denote the energy of $E_c(\tau_i)$, and it can be also balanced by random thermal vibration ($\tau_i$-scale) energy of $kT_g$ for general unexcited particles. That is

$$E_c(\tau_i) = kT_g°(\tau_i) = kT_g(\tau_i) = 20/3\,\varepsilon_0(\tau_i) \tag{10}$$

The numerical value of $T_g°(\tau_i) = T_g°$ that can be called the *fixed point* of clusters from small ($\sigma_1$, $\tau_1$ scale) to large ($\sigma_8$, $\tau_8$ scale) in inverse cascade at GT, it is independent of GT temperature $T_g$. $T_g(\tau_8) = T_g$ is traditionally accepted as GT temperature, which is obtained by slow heating rate. Therefore, $E_c$ is a measurable by experiments. From (2), numerical relationship is:

$$kT_g° = kT_2 + \varepsilon_0 \tag{11}$$

**(d) Geometric frustration and high-density percolation.** It can be seen that **Eq.9** is based on the result of corrected $L_8$ value and this correction from $L_7$ to $L_8$ in **Figure 5** also can be regard as the geometric frustration effect, which appearance corresponds to the GT [2, 3, 8] This also indicates that the percolation at GT belongs to the *high-density percolation* [1]. The filling factor of instantaneous polarized dipole $a_0$ in **Figure 1(d)** is also 1.

### 3.5. Activation Energy to Break Solid Lattice.

**Figure 5** has given out the number of all IE states at GT: the 320 different IE states. The numerical value is from

$$\sum_{i=1}^{8} L_i + 8 = 320 \tag{12}$$

$$= 8 \text{ (spatial evolution)} \times 8 \text{ (temporal evolution)} \times$$
$$5 \text{ (5particles in 5-particle cooperative excited)}$$

The 8 evolution IE states together with the 4 IEs on a reference particle $a_0$ will evolve a new first order of IE loop-flow being of 12 IEs when 8th order cluster appears.

The energy summation of the 320 different IE states is named the *cooperative orientation activation energy* to break solid lattice, denoted as $\Delta E_{co}$. The reason to call it activation energy is that although in microscopical, the energy to break solid lattice is seemly only $kT_g$, $< \Delta E_{co}$, in microcosmic, the energy of 320 IE space-time states in every solid domain is needed. Thus,

$$\Delta E_{co} = 320\Delta\varepsilon = 40\varepsilon_0 \tag{13}$$

$$E_c = kT_g° = 1/6\,\Delta E_{co} \tag{14}$$

Using $T_g°(\tau_8) = T_g$, thus,

$$KT_g = 1/6 \Delta E_{co} \qquad (15)$$

$\Delta E_{co}$ can be measured on melt high-speed spinning-line. In fact, the author [4] had obtained the experimental data of $\Delta E_{co}$ for polyethylene terephthalate (PET) by using the on-line measuring on the stretch orientation zone during melt high-speed spinning. The experimental result shows $kT_g$ (for PET) $\approx 1/6\Delta E_{co}$ (for PET), and $\Delta E_{co} = 2035k$. Thus, for PET, the extra IE energy $\Delta\varepsilon$

$$\Delta\varepsilon = 6.4\,k\,(= 5.5 \times 10^{-4}\text{ eV}) \qquad (16)$$

Moreover, $\varepsilon_0 = 8\Delta\varepsilon = 51k \approx 4.4 \times 10^{-3}$ eV.

$\varepsilon_0$ is the energy of one DoF of $V_8$-loop. This value is consistent with the experimental results ($4 \sim 12 \times 10^{-3}$ eV) for Boson peak measured by high-resolution inelastic neutron scattering [12].

### 3.6. Free Volume

The classical free volume ideas have been questioned because of the misfits with pressure effects, but here the pressure effects should be indeed minor [2]. It is interesting in our model that there is no so-called classical free volume with an atom (molecule) migrating in system. Except the 8th order of 2D clusters, in the $i$-th order of clusters, all $(i-1)$-th order of 2D clusters *are compacted* to form $i$-th order clusters, and the extra volumes of the IEs of all $(i-1)$-th order clusters vanish and reappear on the interfaces of $i$-th order. Thus, when the central 8th order loop appears and vanish in cascade, the extra vacancy volumes of the IEs on the loop suddenly form vacancy volumes of 5 cavity sites in the central of the loop, in **Figure 5**, in order to induce the first order of cluster delocalization. The definition of free volume is modified by the cooperatively appearing 5 cavities, i.e., using clusters rather than atoms, as same as that proposed by de Gennes [2]. The modified free volume idea is still true, because the action of *pressure should be only through the same percolation field generating cavity volume*; and the extra volume energy in the 8 orders of loop-flows in **Figure 5** should be balanced with the external pressure or tension. Each cavity is also an orientation vector as same as the direction of excited domain. Since all excited domains are in random orienting at GT, 5 statistically isotropic 3D cavities (so-called classical free volume) thus are obtained by per 200 statistically isotropic 3D particles. In other words, the free volume fraction: $5/200 = 0.025$ can be directly and explicitly obtained, which is in accordance with the experimental results for flexible polymer.

### 3.7. Mode-Coupling and Icosahedral Directional Ordering

In the inverse cascade at GT, if each IE appears in the way of one arrow after another according to the appearance probability as mentioned in **Figures 2–5**, the probability is rather low. In fact, the more probable situation is the evolutive mode of self-similar 2body-3body cluster coupling in the *soft* matrix of IE loop-flows in **Figure 6**.

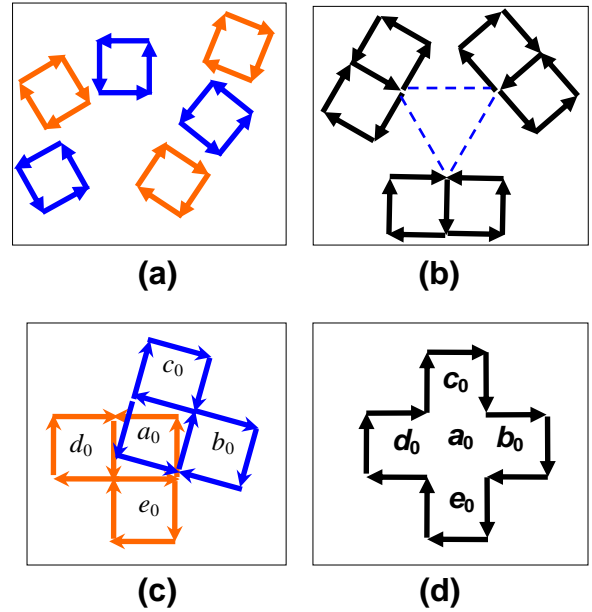

**Figure 6.** The soft matrix of self-similar 2body-3body cluster coupling.

In **Figure 6**, the $i$-th order of IE loop-flows can be simply denoted as some squares with IE $\Delta\varepsilon(\tau_i)$ in **(a)**, each of which is self-similar with **Figure 1(d)**. **Figure 6(a)** first evolves into three 2-body interaction in **(b)**, then into two three-body interactions in **(c)**, named as $i$-th order of 2body-3body cluster coupling in 5-particle cooperative excited field. Finally into $(i+1)$-th order of IE loop-flow in **(d)** and cooperatively eliminate IEs inside the loop-flow (static screen effect) in order to form 8 order 2D IE loop-flows and clusters slightly more compact than the matrix.

The mode of 2body-3body cluster coupling is in statistically advantage (soft matter concept): e.g., per 4 IEs needs 5 instantaneous polarized ions in **Figure 1(d)** and **Figure 6(a)**. The parameter of *polarized ion number-density of per IE* (the density of particles occupy additional vacancy volume in IEs) in soft matrix, $D_{ion}(\tau_0) = 5/4 = 1.25$. However, 21 IEs needs 24 polarized ions, the ion number-density $D_{ion} = 24/21 \approx 1.14$ in **Figure 6(b)**, large than $D_{ion}(\tau_0 \to \tau_1) = 22/20 = 1.1$ in **Figure 6(c)**, each evolutive 3body cluster only needs the excitation energy of 10 IEs. In **Figure 6(c)**, the two clusters (two 3body clusters excited at different times need 20 IEs) move randomly and they are not in superposition. Once they are in superposition, **Figure 6(c)** immediately mutates into **Figure 6(d)** to form first order of uncompacted cluster, $D_{ion}(\tau_1) = 17/16 \approx 1.06$ in **Figure 6(d)**. Finally, the IEs inside the loop-flow disappear to form hard-cluster. The parameter $D_{ion}$ will be less and less in inverse cascade with cluster augment.

The maximum value of the polarized ion number density $D_{ion}(\tau_8) = 200/320 = 5/8$, in 8th order 2D cluster in **Figure 5**, which indicates the single-ion polarization is slow process formed IE closed loop, **Figure 1(d)**, corresponding the *maximum static electricity repulsive state* in polarized field. In addition, there are 60 +$z$-axial hopping ions and 60 −$z$-axial hopping ions in 136 $z$-axial excited particles in **Figure 5**. The 120 ions form 60 pairs of *antiparallel ion-pairs*. So, the *polarized ion pair number density* in IE soft matrix, $D_{pair}(\tau_8) = 120/320 = 3/8$, which indicates the polarization of a pair of ion is fast process, **Figure 1(b)** in IE. It is an important relation equation has been fined that $D_{ion} + D_{pair} = 1$. This implicates the polarized structure of classical boson peak in local polarized filed. Especially, the number of 60 pairs of antiparallel ions inside 8th order loop-flow equals the number of the 60 pairs of synchronal polarized electron charges on the loop-flow. Therefore, $z$-axial slower randomly non-integrable kinetic energy of antiparallel ion-pairs inside 8th order loop-flow also balances the faster induced potential energy on the IE loop-flow, which is a kind of *mode-coupling scheme* between the fast-motion induced charge- electron coupling pairs and the slow-motion delocalized ion pairs.

Moreover, there is still 16 (= 136 − 120) +$z$-axial remainder ions inside the 8th order of loop-flow, which can offer a static charge repulsive force to facilitate the first excited +$z$-axial ion $a_0$ delocalization in **Figure 5**. Since the 16 +$z$-axial ions to drive ion $a_0$ to leave its coordinates needs 320 IEs, 16/320 = 1/20, this leads to each directional delocalization *uncompacted* particle needs 20 $z$-directional excited interfaces with $\tau_8$ in its 8th order of loop-flow. On the other hand, disappearing the 8th order of loop-flow of reference $a_0$ particle also needs its 4 concomitant excited particle, $b_0$, $c_0$, $d_0$, and $e_0$, one by one finish respective 8th order of loop-flows in inverse cascade and delocalize along −$z$-axial. This leads to 5 *compacted* conformational rearrangement particles in cascade precisely need 20 $z$-directional excited interfaces *disappearing* with $\tau_8$. That is another explanation for icosahedral directional ordering at the GT.

Distinctly, the mode-coupling scheme based on mosaic *geometric* structure adopted in this paper is different from the mathematical expressions in the current mode-coupling theory of GT. However, the spirit in both mode-coupling schemes is the same that deals with the coupling of fast-slow relaxation modes and two density modes in structure rearrangements at the GT. In our scheme, we focus on finding out the three direction non-integrable energies $kT_2°$, $kT_g°$ and $kT_m°$ *existing in the coordinate invalidation* from $i$-th order clusters to ($i$+1)-th order in solid-to-liquid transition, whatever the time complications of anharmonic frequencies may be. The mode-coupling trick is that the relaxation time complications have been *beforehand reduced* and replaced by the long time Brownian directional regressions with the $2\pi$ closed loop-flows of IEs in **Figure 5**.

## 4. Conclusions

It is proposed that the soft matrix surrounding clusters is 2D soft nano-scale 8 orders of IE loop-flows; the filling factor of 2D lattices is 1. The key of the theory is that the random thermo-motion kinetic energy, from glass transition to melt transition, should be balanced by the non-integrable random regression vibration energy in random systems. That is, entire molecule-cluster delocalization energy (also the GT energy) origins form the maximum order potential energy in random systems. This 2D structure corresponds to the appearance of boson peak and geometric frustration at the GT, and can directly deduce three non-integrable directional energies: $kT_2°$, $kT_g°$, and $kT_m°$ along one –dimensional direction, respectively corresponding Gibbs temperature $T_2$, glass transition temperature $T_g$, and melt transition temperature $T_m$.

## 5. Acknowledgements

The author is grateful to all colleagues he had the pleasure to collaborate and interact, especially when he found the fundamental physics origin for the orientation activation energy obtained experimentally on melt high-speed spinning-line. In particular, the author would like to thank, in random order, Yuan Tseh Lee and Sheng Hsien Lin of Academia Sinica (Taiwan), Yun Huang of Beijing University, Da-Cheng Wu of Sichuan University for useful discussions.